\documentclass[twocolumn,showpacs,showkeys]{revtex4}
\usepackage{graphicx}

\newcommand{\bastar}{\begin{eqnarray*}}
\newcommand{\eastar}{\end{eqnarray*}}
\newskip\humongous \humongous=0pt plus 1000pt minus 1000pt

\relax
\newcommand{\be}{\begin{equation}}
\newcommand{\ee}{\end{equation}}
\newcommand{\bea}{\begin{eqnarray}}
\newcommand{\eea}{\end{eqnarray}}
\newcommand{\X}{{\vec X}}
\newcommand{\pro}{\partial}
\newcommand{\n}{\hat n}
\newcommand{\oneg}{\displaystyle\frac{1}{g}}

\newcommand{\D}{{\hat D}}

\newcommand{\valpha}{{\vec \alpha}}

\newcommand{\dfrac}{\displaystyle\frac}
\newcommand{\ba}{\begin{array}}
\newcommand{\ea}{\end{array}}

\newcommand{\nn}{\nonumber}
\newcommand{\hn}{\hat n}
\begin{document}
\title{Knot Topology of QCD Vacuum}
\author{Y. M. Cho}
\email{ymcho@yongmin.snu.ac.kr}
\affiliation{C. N. Yang Institute
for Theoretical Physics, \\
State University of New York, Stony Brook, New York 11794, USA \\
and \\
School of Physics, College of Natural Sciences, Seoul National University,
Seoul 151-742, Korea\\}
\begin{abstract}
We show that one can express the knot equation of Skyrme theory
completely in terms of the vacuum potential of $SU(2)$ QCD,
in such a way that the equation is viewed as a generalized Lorentz 
gauge condition which selects one vacuum for each class of 
topologically equivalent vacua.  
From this we show that there are three ways to 
describe the QCD vacuum (and thus the knot), by a non-linear sigma 
field, a complex vector field, or by an Abelian gauge potential. 
This tells that the QCD vacuum can be classified by 
an Abelian gauge potential with an Abelian Chern-Simon index.
 
\end{abstract}
\pacs{03.75.Fi, 05.30.Jp, 67.40.Vs, 74.72.-h}
\keywords{knot topology of QCD vacuum, QCD vacuum as knot, 
classification of QCD vacuum by Abelian Chern-Simon index}
\maketitle

The non-Abelian gauge theory has been well known to have a 
non-trivial topology. In particular it has 
infinitely many topologically distinct vacua which can be 
connected by tvacuum tunneling through the instantons \cite{bpst,cho79}. 
The existence of topologically distinct vacua and the vacuum 
tunneling has played a very important role in quantum chromodynamics 
(QCD) \cite{thooft,peccei}. 
In a totally independent development the Skyrme theory has been shown
to admit a topologically stable knot which can be interpreted as 
a twisted magnetic vortex ring made of helical baby 
skyrmion \cite{skyr,fadd1,cho01prl}. 
And very interestingly, this knot 
is shown to describe the topologically distinct QCD 
vacua \cite{cho01,baal}. 

This is puzzling because the knot 
is a physical object which carries a nonvanishing energy.
So it appears strange that the knot can be related to a QCD vacuum.
On the other hand this is understandable since the Skyrme theory
is closely related to QCD, and both the knot and 
the QCD vacuum are described by the same topology 
$\pi_3(S^3)$. Under this circumstance one need to know 
in exactly what sense the QCD vacuum can be identified as the knot.
Since there exists one knot solution for each topological quantum number
(up to the trivial space-time translation and the global $SU(2)$
rotation), one might suspect that the knot equation 
could be viewed as a gauge condition
for the topologically equivalent vacua. In fact it has been
suggested that the knot equation can be viewed as 
a non-local gauge condition which describes
the maximal Abelian gauge in $SU(2)$ QCD \cite{baal}. 
{\it The purpose of this Letter is to show that the knot equation 
is nothing but a generalized Lorentz gauge condition
which selects one representative vacuum for each class
of topologically equivalent QCD vacua. This allows us to interpret
the knot as a complex vector field which couples to
an Abelian gauge field, 
and the knot equation as an Abelian gauge condition 
for the complex vector field.} 
We first obtain a most general
expression of the vacuum, and write the knot equation completely 
in terms of the vacuum potential. With this
we prove that the knot equation is nothing but a generalized
Lorentz gauge condition of the QCD vacuum. From this we show that 
the knot equation can be viewed as an Abelian gauge 
condition for a complex vector field. Moreover, we show that 
this complex vector field is uniquely determined by 
the Abelian gauge potential. This allows a new interpretation
of the knot, the knot as a complex vector field or
an Abelian gauge potential. As importantly, this tells that one
can classify the topologically different QCD vacua by
an Abelian Chern-Simon index.

A best way to describe the QCD vacuum is 
to introduce a local orthonormal frame in 
the non-Abelian group space and obtain a potential which
parallelizes the local orthonormal frame.
Consider the $SU(2)$ QCD and let $\hat n_i~(i=1,2,3)$ be a 
right-handed local orthonormal frame. A vacuum potential must be the one
which parallelizes the local orthonormal frame. Imposing 
the condition to the gauge potential $\vec A_\mu$
\bea
D_\mu \n_i = (\pro_\mu +g \vec A_\mu \times)~\n_i = 0,
~~~(i=1,2,3)
\label{vcon}
\eea
we obtain a most general vacuum potential
\bea
&\hat \Omega_\mu = - C_\mu \n - \dfrac{1}{g} \hn \times \pro_\mu \n 
= - C_\mu^k~\hn_k, \nn\\
&\dfrac{1}{g} \hn \times \pro_\mu \n
= C_\mu^1~\hn_1 + C_\mu^2~\hn_2, \nn\\
&C_\mu^k = -\dfrac{1}{2g} \epsilon_{ij}^{~~k} (\n_i \cdot \pro_\mu \n_j),
\label{vac}
\eea
where $\hn$ is $\hn_3$ and $C_\mu$ is $C_\mu^3$. 
One can easily check that $\hat \Omega_\mu$ describes a vacuum
\bea
&\hat \Omega_{\mu\nu} = \pro_\mu \hat \Omega_\nu
-\pro_\nu \hat \Omega_\mu + g \hat \Omega_\mu \times \hat \Omega_\nu \nn\\
&=-(\pro_\mu C_\nu^k -\pro_\nu C_\mu^k 
+ g \epsilon_{ij}^{~~k} C_\mu^i C_\nu^j)~\n_k = 0.
\label{vacf}
\eea
This tells that both $\hat \Omega_\mu$ and 
$(C_\mu^1,C_\mu^2,C_\mu^3)$ describe a QCD vacuum. 
Obviously they are gauge equivalent.
Notice that the vacuum is essentially fixed by
$\hn$, because $\hn_1$ and $\hn_2$ are uniquely determined by 
$\hn$ up to a $U(1)$ gauge transformation which leaves $\hn$
invariant. 

A nice feature of (\ref{vac}) is that the topological 
character of the vacuum is naturally inscribed in it. The topology 
of the $SU(2)$ QCD vacuum has been described by the non-trivial
mapping $\pi_3(S^3)$ from the (compactified) three-dimensional space 
$S^3$ to the group space $S^3$. But $\hn$ can also describes
the vacuum topology because it defines the mapping $\pi_3(S^2)$
which can be transformed to $\pi_3(S^3)$ through 
the Hopf fibering \cite{cho79}. So
one can naturally classify the vacuum topology by $\hn$,
which is manifest in (\ref{vac}).

With
\bea
\hn = \Bigg(\matrix{\sin{\alpha}\cos{\beta} \cr
\sin{\alpha}\sin{\beta} \cr \cos{\alpha}}\Bigg),
\label{n}
\eea
one may choose
\bea
\hn_1 = \Bigg(\matrix{\cos{\alpha}\cos{\beta} \cr
\cos{\alpha}\sin{\beta} \cr -\sin{\alpha}}\Bigg),
~~~~~\hn_2 = \Bigg(\matrix{-\sin{\beta} \cr
\cos{\beta} \cr 0}\Bigg),
\eea
so that one has
\bea
&C_\mu^1= - \oneg \sin{\alpha} \pro_\mu \beta,
~~~~~C_\mu^2 = \oneg \pro_\mu \alpha, \nn\\
&C_\mu = \oneg \cos{\alpha} \pro_\mu \beta.
\eea
Of course they are uniquely determined up to the $U(1)$ gauge 
transformation which leaves $\hn$ invariant.
Notice that, when $\hn$ becomes the unit radial vector $\hat r$,
$C_\mu$ describes the well-known Dirac's monopole potential.
But when $\hn$ is smooth everywhere, it describes a vacuum.

The vacuum (\ref{vac}) is obtained by three conditions given by
(\ref{vcon}). Suppose we impose only one condition 
\bea
D_\mu \n = 0.~~~~~(\hn=\hn_3)
\label{ccon}
\eea
This singles out the restricted potential which defines 
the restricted gauge theory \cite{cho80,cho81}
\bea
&\hat A_\mu =A_\mu \n - \oneg \n \times \pro_\mu \n 
=\hat \Omega + (A_\mu + C_\mu) \hn,
\label{rpot}
\eea
where $A_\mu = \n\cdot \vec A_\mu$ is the chromoelectric potential. 
This tells that the two extra conditions (for $i=1,2$) 
in (\ref{vcon}) uniquely determines $A_\mu$ to be
\bea
A_\mu=-C_\mu. 
\eea
Indeed with this (\ref{rpot}) becomes (\ref{vac}), 
which tells that the restricted QCD has exactly the
same multiple vacua. Furthermore, in the absence of (\ref{ccon}),
one can express the most general $SU(2)$ gauge potential 
$\vec A_\mu$ by \cite{cho80,cho81}
\bea
\vec A_\mu = \hat A_\mu + \vec X_\mu,~~~~~(\hn \cdot \vec X_\mu=0)
\label{cdec}
\eea
where $\vec X_\mu$ is a gauge covariant vector field.
This is because under the infinitesimal gauge transformation 
\bea
\delta \hat n = - \valpha \times \hn,
\label{gt}
\eea
one has
\bea
&\delta A_\mu = \oneg \n \cdot \pro_\mu \valpha,
~~~~~\delta C_\mu = - \delta A_\mu, \nn\\
&\delta \hat A_\mu = \oneg \D_\mu \valpha,  
~~~~~\delta \X_\mu = - \alpha \times \X_\mu,
\eea
where $\valpha$ is the infinitesimal gauge parameter.
This means that one can interpret QCD as a restricted gauge theory
which has a gauge covariant valence gluon $\X_\mu$ as
the colored source \cite{cho80,cho81}.
The importance of the decompositions is that they are gauge
independent. Once $\hn$ is given the decomposotion follows
automatically, independent of the choice of a gauge.

The above analysis shows that $\hat A_\mu$ by itself describes
an $SU(2)$ connection which
enjoys the full non-Abelian gauge degrees of freedom. 
More importantly, it has a dual structure \cite{cho80,cho81}
\bea
&\hat F_{\mu\nu} = (F_{\mu\nu}+H_{\mu\nu}) \hn, \nn\\
&F_{\mu\nu}=\partial_\mu A_{\nu}-\partial_{\nu}A_\mu, \nn\\
&H_{\mu\nu}= -\dfrac{1}{g} \hn \cdot \pro_\mu \n \times \pro_\nu \n
=\partial_\mu C_\nu-\partial_\nu C_\mu.
\eea
This tells that $C_\mu$ in (\ref{vac}) is nothing but 
the chromomagnetic potential of the field strength $H_{\mu\nu}$ 
(Since $H_{\mu\nu}$ forms a closed two-form, it
admits a potential). Moreover $A_\mu$ and $C_\mu$ transform
equally but oppositely under the gauge transformation.
In particular, the Abelian gauge group
which leaves $\hn$ invariant acts on both $A_\mu$ and $C_\mu$.
This shows that the restricted QCD has a manifest 
electric-magnetic duality \cite{cho80,cho81}.

Now, let's review the knot in Skyrme theory \cite{skyr,cho01prl}.
Let $\omega$ and $\hat n$ (with ${\hat n}^2 = 1$) be the Skyrme
field and the non-linear sigma field, and let
\bea
&U = \exp (\dfrac{\omega}{2i} \vec \sigma \cdot \hat n)
= \cos \dfrac{\omega}{2} - i (\vec \sigma \cdot \hat n)
\sin \dfrac{\omega}{2}, \nn\\
&L_\mu = U\partial_\mu U^{\dagger}.
\label{su2}
\eea
With this one can write the Skyrme Lagrangian as 
\bea
&{\cal L}_S = \dfrac{\mu^2}{4} {\rm tr} ~L_\mu^2
+ \dfrac{1}{32}
{\rm tr} \left( \left[ L_\mu, L_\nu \right] \right)^2 \nn\\
&= -\dfrac{g^2}{4} (1-\sigma^2)^2 \hat H_{\mu\nu}^2
-\mu^2 \dfrac{g^2}{2} (1-\sigma^2) \hat C_\mu^2 \nn\\
&-\dfrac{\mu^2}{2} \dfrac{(\partial_\mu \sigma)^2}{1-\sigma^2}
-\dfrac{g^2}{4} (\partial_\mu
\sigma \hat C_\nu - \partial_\nu \sigma \hat C_\mu)^2,
\label{slag}
\eea
where 
\bea
&\sigma= \cos \dfrac{\omega}{2}, \nn\\
&\hat H_{\mu\nu} = \partial_\mu \hat C_\nu - \partial_\nu \hat C_\mu
+ g \hat C_\mu \times \hat C_\nu = H_{\mu\nu} \hn, \nn\\
&\hat C_\mu = -\dfrac{1}{g} \hn \times \partial_\mu \hn. \nn
\eea
The Lagrangian has an obvious global $SU(2)$ symmetry,
but it also has a (hidden) $U(1)$ gauge symmetry \cite{cho01prl}.
This is because the invariant subgroup of $\hn$ can be viewed as 
a $U(1)$ gauge group. Notice that $\hat C_\mu$ is nothing
but the magnetic part of the restricted potential
(\ref{rpot}). This provides the crucial link between QCD and 
Skyrme theory. From this link one can argue that the Skyrme theory 
is a theory of monopole which describes the chromomagnetic 
dynamics of QCD \cite{cho01prl}. 

It is well-known that $\sigma=0$ is a classical solution 
of (\ref{slag}), independent of $\hn$. When $\sigma=0$ 
the Skyrme Lagrangian is reduced to 
\bea
&{\cal L}_{S} \rightarrow -\dfrac{1}{4} \hat H_{\mu\nu}^2
- \dfrac{\mu^2}{2} \hat C_\mu^2, 
\label{qcdlag}
\eea
whose equation is given by 
\bea
&\dfrac{1}{g} \hn \times \partial^2 \hn
- \dfrac{1}{\mu^2} (\partial_\mu H_{\mu\nu})
\partial_\nu \hn = 0.
\label{keq1}
\eea
It is this equation that allows the monopole,
the baby skyrmion, and the knot in 
Skyrme theory \cite{cho01prl,cho01}. 

With (\ref{n}) the knot equation is written as
\bea
& \pro^2 \alpha - \sin{\alpha} \cos{\alpha} 
(\pro_\mu \beta)^2 
- \dfrac{g}{\mu^2} \sin{\alpha} (\pro_\mu H_{\mu\nu}) 
\pro_\nu \beta =0, \nn\\
&\sin{\alpha} \pro^2 \beta + 2 \cos{\alpha} 
(\pro_\mu \alpha \pro_\mu \beta) 
+ \dfrac{g}{\mu^2} (\pro_\mu H_{\mu\nu}) \pro_\nu \alpha \nn\\
&=0,
\label{keq2}
\eea
where
\bea
H_{\mu\nu} = -\oneg \sin{\alpha} (\pro_\mu \alpha \pro_\nu \beta
- \pro_\nu \alpha \pro_\mu \beta). \nn
\eea
But this can neatly be expressed by the vacuum potential $C_\mu^i$.
To see this, notice that the knot equation (\ref{keq1})
can be understood as a conservation equation of an $SU(2)$ current
$\vec j_\mu$
\bea
&\vec j_\mu = \oneg \hn \times \pro_\mu \hn 
-\dfrac{1}{\mu^2} H_{\mu\nu} \pro_\nu \hn \nn\\
&= (C_\mu^1 - \dfrac{1}{\mu^2} H_{\mu\nu} C_\nu^2)~\hn_1
+ (C_\mu^2 + \dfrac{1}{\mu^2} H_{\mu\nu} C_\nu^1)~\hn_2, \nn\\
&\pro_\mu \vec j_\mu =0.
\label{keq3}
\eea
The origin of this conserved current, of course,
is the global $SU(2)$ symmetry of the Skyrme Lagrangian
(\ref{slag}). From (\ref{keq3}) 
one can express the knot equation by
\bea
&\pro_\mu C_\mu^1 - g C_\mu C_\mu^2
- \dfrac{g}{\mu^2} (\pro_\mu H_{\mu\nu})~C_\nu^2 =0, \nn\\
&\pro_\mu C_\mu^2 + g C_\mu C_\mu^1 + \dfrac{g}{\mu^2}
(\pro_\mu H_{\mu\nu})~C_\nu^1 =0.
\label{keq4}
\eea
With
\bea
\omega_\mu = \dfrac{C_\mu^1+iC_\mu^2}{\sqrt 2},
\label{compp}
\eea
this can be put into a single complex equation
\bea
&\bar D_\mu \omega_\mu =0, \nn\\
&\bar D_\mu = \pro_\mu + i g (C_\mu + \dfrac{1}{\mu^2} 
\pro_\alpha H_{\alpha\mu}).
\label{compkeq}
\eea
This tells that the knot equation (\ref{keq1}) can be expressed
completely in terms of the QCD vacuum potential, as an Abelian 
gauge condition for the complex vector field $\omega_\mu$. 
In this form the $U(1)$ gauge symmetry of the knot equation 
(and the Skyrme theory) becomes manifest \cite{cho01prl}.

Now let us go back to QCD, and consider the following 
constraint equation for the vacuum
\bea
\bar D_\mu \hat \Omega_\mu =0,
\label{vgcon1}
\eea
where now
\bea
\bar D_\mu = \pro_\mu + g (C_\mu +
\dfrac{1}{\mu^2} \pro_\alpha H_{\alpha\mu}) \hn \times. \nn
\eea
This is equivalent to
\bea
&\bar D_\mu \omega_\mu =0,
~~~~~\pro_\mu C_\mu =0.
\label{vgcon2}
\eea
This tells that the equation (\ref{vgcon1}) not only
describes the knot, but also fixes the $U(1)$ gauge degree 
of the knot equation. This proves that the knot equation can
be interpreted as a generalized Lorentz gauge condition of
the QCD vacuum which selects one vacuum for each topologically 
equivalent class of vacua.

The knot equation (\ref{compkeq}) contains both $\omega_\mu$
and $C_\mu$. But they are not independent.
To see this, notice that the vacumm
condition (\ref{vacf}) tells that
\bea
&D_\mu \omega_\nu - D_\nu \omega_\mu = 0, \nn\\
&H_{\mu\nu} = \pro_\mu C_\nu-\pro_\nu C_\mu 
= ig(\omega_\mu^* \omega_\nu - \omega_\nu^* \omega_\mu), 
\label{vcon2}
\eea
where
\bea
D_\mu = \pro_\mu + ig C_\mu. \nn
\eea
So $\omega_\mu$ and $C_\mu$ are determined by 
each other. This tells that $\omega_\mu$
alone can describe the knot. Equivalently,
this means that the knot can also
be described by an Abelian gauge potential $C_\mu$.
So we have three different ways to describe the knot and thus
the QCD vacuum, by $\hn$, $\omega_\mu$, and $C_\mu$.

With (\ref{vcon2}) we have 
\bea
\bar D_\mu \omega_\mu = D_\mu \omega_\mu 
+i \dfrac{g}{\mu^2} \big[\pro_\mu (H_{\mu\nu} \omega_\nu) 
+ig H_{\mu\nu} C_\mu \omega_\nu \big],
\eea
so that we can simplify (\ref{compkeq}) to
\bea
&D_\mu \bar \omega_\mu = 0, \nn\\
&\bar \omega_\mu = \omega_\mu 
+ i \dfrac{g}{\mu^2} H_{\mu\nu} \omega_\nu.
\label{compkeq2}
\eea
This tells that the knot equation can be expressed as a 
covariant Lorentz gauge condition of the complex 
vector field $\bar \omega_\mu$. 

The knot quantum number is given by the Abelian Chern-Simon
index of the magnetic potential $C_\mu$ \cite{fadd1,cho01prl},
\bea
&Q = \dfrac{g^2}{32\pi^2} \int \epsilon_{ijk} C_i H_{jk} d^3x,
\label{kqn}
\eea
which describes the non-trivial topology $\pi_3(S^2)$
defined by $\hn$. The preimage of the mapping from
the compactified space $S^3$ to the target space $S^2$
defined by $\hn$ forms a closed circle, and any two
preimages of the mapping are linked together when
$\pi_3(S^2)$ is non-trivial. This linking number  
is given by the Chern-Simon index.
Obviously our analysis tells that exactly the same description 
applies to the QCD vacuum.
In particular, this means that the QCD vacuum can also
be classified by an Abelian gauge potential 
with the Abelian Chern-Simon index \cite{cho79,cho01}.

Conversely, with (\ref{vcon2}) we can transform the knot
quantum number (\ref{kqn}) to a non-Abelian form
\bea
&Q = \dfrac{g^2}{32\pi^2} \int \epsilon_{ijk} C_i H_{jk} d^3x \nn\\
&=-\dfrac{g^3}{96\pi^2} \int \epsilon_{abc} 
\epsilon_{ijk} C_i^a C_j^b C_k^c d^3x,
\label{nakqn}
\eea
which proves that the knot quantum number can also
be expressed by a non-Abelian Chern-Simon index.
More significantly, this tells that the Abelian Chern-Simon index
is actually identical to the non-Abelian Chern-Simon index.
They have been thought to be two different things, but our 
analysis tells that they are one and the same thing which
can be transformed to each other through the
vacuum condition (\ref{vcon2}).

The fact that the knot can be described by an Abelian gauge potential
$C_\mu$ raises a totally unexpected and very interesting 
possibility that, under a proper circumstance, 
one could create the knot in a condensed matter. 
Indeed it has been conjectured that
a superconducting knot could exist in the ordinary 
superconductor \cite{chocm3}. It has long been assumed
that this is impossible, because the Abelian gauge theory
is thought to be too trivial to allow the knot topology \cite{huang}.   
Our analysis tells that this is not true. There exists a well-defined
knot topology in the Abelian gauge theory.

Our analysis could have important applications in QCD. 
For example, the decomposition (\ref{cdec}) plays
an important role in the discussion of the Abelian domonance
and the confinement of color in QCD \cite{cho02,cho04}. 
Moreover it plays a crucial role for us to study the geometry 
of the principal fiber bundle, in particular the Deligne cohomology 
of the non-Abelian gauge theory \cite{cho75,zucc}.
Further interesting applications of our analysis to QCD
will be published elsewhere \cite{cho1}.

{\bf ACKNOWLEDGEMENT}

~~~The author thanks Professor C. N. Yang for the
illuminating discussions, and G. Sterman for the kind hospitality
during his visit to Institute for Theoretical Physics. The work is
supported in part by the ABRL Program (Grant
R14-2003-012-01002-0) of Korea Science and Engineering
Foundation, and by the BK21 project
of Ministry of Education.


\begin{thebibliography}{99}
\bibitem{bpst}A. Belavin, A. Polyakov, A. Schwartz, and Y. Tyupkin,
Phys. Lett. {\bf B59}, 85 (1975); C. Callan, R. Dashen, and D. Gross,
Phys. Lett. {\bf B63}, 334 (1976); R. Jackiw and C. Rebbi, Phys. Rev. Lett.
{\bf 37}, 172 (1976).
\bibitem{cho79}Y. M. Cho, Phys. Lett. {\bf B81}, 25 (1979).
\bibitem{thooft} G. 't Hooft, Phys. Rev. Lett. {\bf 37}, 8 (1976);
Phys. Rev. {\bf D14}, 3432 (1976).
\bibitem{peccei} R. Peccei and H. Queen, Phys. Rev. Lett. {\bf 38}, 
1440 (1977); S. Weinberg, Phys. Rev. Lett. {\bf 40}, 223 (1978);
F. Wilczek, Phys. Rev. Lett. {\bf 40}, 279 (1978).
\bibitem{skyr}T. Skyrme, Proc. Roy. Soc. (London) {\bf 260}, 127
(1961); {\bf 262}, 237 (1961); Nucl. Phys. {\bf 31}, 556 (1962).
\bibitem{fadd1} L. Faddeev and A. Niemi, Nature {\bf 387}, 58 (1997);
J. Gladikowski and M. Hellmund, Phys. Rev. {\bf D56}, 5194 (1997);
R. Battye and P. Sutcliffe, Phys. Rev. Lett. {\bf 81}, 4798 (1998).
\bibitem{cho01prl} Y. M. Cho, Phys. Rev. Lett. {\bf 87}, 252001 (2001);
Y. M. Cho and B. S. Park, hep-th/0404181.
\bibitem{cho01}W. S. Bae, Y. M. Cho, and S. W. Kimm, Phys. Rev. 
{\bf D65}, 025005 (2001).
\bibitem{baal} P. van Baal and A. Wipf, Phys. Lett. {\bf B515},
181 (2001).
\bibitem{cho80} Y. M. Cho, Phys. Rev. {\bf D21}, 1080 (1980);
Phys. Rev. {\bf D62}, 074009 (2000);
L. Faddeev and A. Niemi, Phys. Rev. Lett.
{\bf 82}, 1624 (1999); Phys. Lett. {\bf B449}, 214 (1999);
{\bf B464}, 90 (1999).
\bibitem{cho81} Y. M. Cho, Phys. Rev. Lett. {\bf 46}, 302 (1981);
Phys. Rev. {\bf D23}, 2415 (1981);
S. Shabanov, Phys. Lett. {\bf B458}, 322 (1999);
H. Gies, Phys. Rev. {\bf D63}, 125023 (2001).
\bibitem{chocm3} Y. M. Cho, cond-mat/0311201.
\bibitem{huang} K. Huang and R. Tipton, Phys. Rev. {\bf D23}, 3050 (1981).
\bibitem{cho02} Y. M. Cho, H. W. Lee, and D. G. Pak,
Phys. Lett. {\bf B 525}, 347 (2002); Y. M. Cho and D. G. Pak,
Phys. Rev. {\bf D65}, 074027 (2002); K. Kondo, hep-th/0404252.
\bibitem{cho04} Y. M. Cho, M. Walker, and D. G. Pak, 
JHEP {\bf 05}, 073 (2004). Y. M. Cho, hep-th/0301103.
\bibitem{cho75} Y. M. Cho, J. Math. Phys. {\bf 16}, 2029 (1975);
Y. M. Cho and P. S. Jang, Phys. Rev. {\bf D12}, 3789 (1975).
\bibitem{zucc} R. Zucchini, hep-th/0306287.
\bibitem{cho1} Y. M. Cho, to be published.
\end{thebibliography}
\end{document}